\documentclass[reprint,aps,pra,superscriptaddress,floatfix]{revtex4-2}
\usepackage{graphicx}
\usepackage{dcolumn}
\usepackage{bm}
\usepackage{float}
\usepackage{amsmath}
\usepackage{amssymb} 
\usepackage[table]{xcolor}
\usepackage{xspace}
\usepackage{xcolor}
\usepackage{color,colortbl}
\usepackage{enumerate}
\usepackage[left=23mm,right=13mm,top=35mm,columnsep=15pt]{geometry} 
\usepackage[english]{babel}
\usepackage[utf8]{inputenc}
\usepackage{cancel}
\usepackage[pagewise]{lineno}
\usepackage{amsthm}
\usepackage{mathtools}
\usepackage{chemformula} 
\usepackage{adjustbox}
\usepackage[T1]{fontenc}
\usepackage{lipsum}
\usepackage{lineno}
\usepackage{cellspace}
\usepackage{multirow,bigstrut}
\usepackage[normalem]{ulem}
\usepackage[update,prepend]{epstopdf}
\usepackage[breaklinks=true,colorlinks=true,linkcolor=blue,urlcolor=blue,citecolor=blue]{hyperref} 
\usepackage[capitalise]{cleveref}
\usepackage{csquotes} 
\usepackage{gensymb}
\usepackage{microtype}

\definecolor{amber(sae/ece)}{rgb}{1.0, 0.49, 0.0}
\definecolor{airforceblue}{rgb}{0.36, 0.54, 0.66}
\definecolor{alizarin}{rgb}{0.82, 0.1, 0.26}
\definecolor{antiquefuchsia}{rgb}{0.57, 0.36, 0.51}
\definecolor{applegreen}{rgb}{0.55, 0.71, 0.0}
\definecolor{blue(pigment)}{rgb}{0.2, 0.2, 0.6}
\definecolor{BrightGray}{rgb}{.94,.94,.94}

\footnotetext{These authors contributed equally to this work}

\usepackage{tikz,xcolor,hyperref}

\definecolor{lime}{HTML}{A6CE39}
\DeclareRobustCommand{\orcidicon}{%
	\begin{tikzpicture}
	\draw[lime, fill=lime] (0,0) 
	circle [radius=0.16] 
	node[white] {{\fontfamily{qag}\selectfont \tiny ID}};
	\draw[white, fill=white] (-0.0625,0.095) 
	circle [radius=0.007];
	\end{tikzpicture}
	\hspace{-2mm}
}

\foreach \x in {A, ..., Z}{%
	\expandafter\xdef\csname orcid\x\endcsname{\noexpand\href{https://orcid.org/\csname orcidauthor\x\endcsname}{\noexpand\orcidicon}}
}

\begin{document}

\title{Tunable ultrafast thermionic emission from femtosecond-laser hot spot on a metal surface: role of laser polarization and angle of incidence}

\author{Mousumi Upadhyay Kahaly$^{\S}$\orcidF{}}
 \affiliation{ELI-ALPS, ELI-HU Non-Profit Ltd., Wolfgang Sandner utca 3., Szeged 6728, Hungary}
 \affiliation{Institute of Physics, University of Szeged, D\'{o}m t\'{e}r 9, H-6720 Szeged, Hungary}

\author{Saibabu Madas$^{\S}$\orcidB{}}
\affiliation{ELI-ALPS, ELI-HU Non-Profit Ltd., Wolfgang Sandner utca 3., Szeged 6728, Hungary}
\affiliation{Institute of Physics, University of Szeged, D\'{o}m t\'{e}r 9, H-6720 Szeged, Hungary}

\author{Boris Mesits}
\affiliation{ELI-ALPS, ELI-HU Non-Profit Ltd., Wolfgang Sandner utca 3., Szeged 6728, Hungary}%
\affiliation{Department of Physics and Astronomy, University of Pittsburgh, Pittsburgh, PA15213, USA}

\author{Subhendu Kahaly\orcidA{}}%
 \email[Email: ]{subhendu.kahaly@eli-alps.hu}
 \affiliation{ELI-ALPS, ELI-HU Non-Profit Ltd., Wolfgang Sandner utca 3., Szeged 6728, Hungary}
 \affiliation{Institute of Physics, University of Szeged, D\'{o}m t\'{e}r 9, H-6720 Szeged, Hungary}
%

\date{\today}
\begin{abstract}
Ultrafast laser induced thermionic emission from metal surfaces has several applications. Here, we investigate the role of laser polarization and angle of incidence on the ultrafast thermionic emission process from laser driven gold coated glass surface. The spatio-temporal evolution of electron and lattice temperatures are obtained using an improved three-dimensional (3D) two-temperature model (TTM) which takes into account the 3D laser pulse profile focused obliquely onto the surface.  The associated thermionic emission features are described through modified Richardson-Dushman equation, including dynamic space charge effects and are included self-consistently in our numerical approach. We show that temperature dependent reflectivity influences laser energy absorption. The resulting peak electron temperature on the metal surface monotonically increases with angle of incidence for P polarization, while for S polarization it shows opposite trend. We observe that thermionic emission duration shows strong dependence on angle of incidence and contrasting polarization dependent behaviour. The duration of thermionic current shows strong correlation to the intrinsic electron-lattice thermalization time, in a fluence regime well below the damage threshold of gold. The observations and insights have important consequences in designing ultrafast thermionic emitters based on a metal based architecture. 
\end{abstract}


\maketitle
\section{Introduction}
\label{sec:intro}

Ultrafast intense laser (typical peak intensity $> 10^{19}~Wcm^{-2}$ and central wavelength $\simeq 800~nm$) interacting with any optical quality material surface instantly ionizes it into a conducting dense plasma medium, in its rising edge. The peak of the femtosecond pulse then interacts with the reflective dynamic plasma surface and the associated nonequilibrium energy transport has wide-ranging applications in plasma optics \cite{Vincenti2014}, attosecond pulse generation \cite{Chopineau2021} and charge particle acceleration \cite{Thvenet2015, DeMarco2023}, etc. At moderate peak intensities ($\sim 10^{9}-10^{14}~Wcm^{-2}$), for ultrashort laser excited surface the nature of the interaction depends on the specific material properties. In case of a metal surface, it undergoes ultrafast thermal evolution which lies at the core of several applications like thermoreflectance, laser induced material damage, laser machining or thermionic emission \cite{Madas2018,UK2017}, to name a few. Under femtosecond laser irradiation, a metal demonstrates complex interplay of electron and lattice dynamics \cite{Rethfeld2004} resulting in time-dependent evolution of temperature of electron and lattice sub-systems \cite{anisimov1974electron,nessler1998,Chen2011}, associated carrier density fluctuation and finally equilibration dynamics. The energy deposition, redistribution and equilibration dynamics can be captured through the thermal evolution of the sub-systems and holds key to many potential applications. 

\par Several applications relevant for next generation electronics are based on metal semiconductor interfaces. These rely on ultrafast thermionic carrier injection from the metal film to the semiconductor layer under conditions when the metal electron temperatures are highly elevated with respect to its lattice temperature \cite{Tomko2020,Keller2022}. On a more fundamental level, the thermal management and ultrafast cooling dynamics of localized hot spots on optically excited metal surface \cite{Sivan2020} or metal nano structures \cite{Siemens2009} on dielectrics is important for investigations on heat transport. In addition, thermionic emission based refrigeration has been a theoretical proposal \cite{ Mahan1994}, that might have potential applications. Recently, experiments on localized, transient thermal excitation of metals have shown signatures of induced surface cooling due to the ensuing emission \cite{Kerse2016,Tomko2022}.

\par The concept of two-temperature model (TTM) as proposed by Anisimov and coworkers \cite{anisimov1974electron} is often used to describe the energy relaxation of excited electrons in metals, effective emission of electrons and overall effect of high-power laser irradiation on a metal surface. Often one-dimensional (1D) \cite{Qiu1994,JKChenJEBeraun2001,Christensen2007,Gurevich2017,D2023} or two-dimensional (2D) TTM \cite{zhang2017numerical} are used to describe the laser-induced thermal evolution in metals. However, such models are unable to describe the complete three-dimensional spatio-temporal variation induced by the finite focal spot size of the ultrashort laser on the metal film. Neither the laser polarization effects nor the effects due to varying angle of incidence are taken into account. Thus, three-dimensional (3D) TTM \cite{Zhang2015,Bresson2020, KhosraviKhorashad2022} is a more suitable description, essential to capture the effective 3D heat deposition by the laser source, and the subsequent thermal evolution of the irradiated thin surface and the volume underneath. 

\par In this work, first, we propose an improved 3D TTM model, where  generalized non-linear heat equation and laser interactions are described in all three spatial directions (\textit{x}, \textit{y}, \textit{z}) to evaluate the spatial and temporal profiles of electron and lattice temperatures when a nanometric gold coated glass substrate is irradiated with a focused Gaussian femtosecond laser pulse. The framework allows us to investigate the effect of oblique incidence excitation, influence of laser polarization and the impact of hot spot due to the finite focal spot size. Then we self-consistently couple 3D TTM with suitable thermionic emission description, and simulate ultrafast laser induced energy deposition and resulting absorption within the gold film, and further elucidate essential space-charge-led modification in thermionic currents. While the electronic temperature, lattice temperature and electron-lattice thermalization time are calculated using the TTM based approach, the characteristics of thermionic emission of gold films under ultrashort laser excitation are investigated using a modified Richardson-Dushman (MRD) equation that includes the dynamic effect of space-charge. Our results reveal that the maximum polarization sensitivity of the thermionic current can be achieved at larger angle of incidence. The investigations unravel a clear dependence of the thermionic current and emission duration on the electron-lattice thermalization dynamics, a correlation not reported in the existing literature. Our analysis suggest ways to tune the duration of thermionic current, which is a significant feature for further applications of thermionic emission. 

\begin{figure*}[th!]
\centering
\vskip -0.45cm 
\includegraphics[width=.70\textwidth]{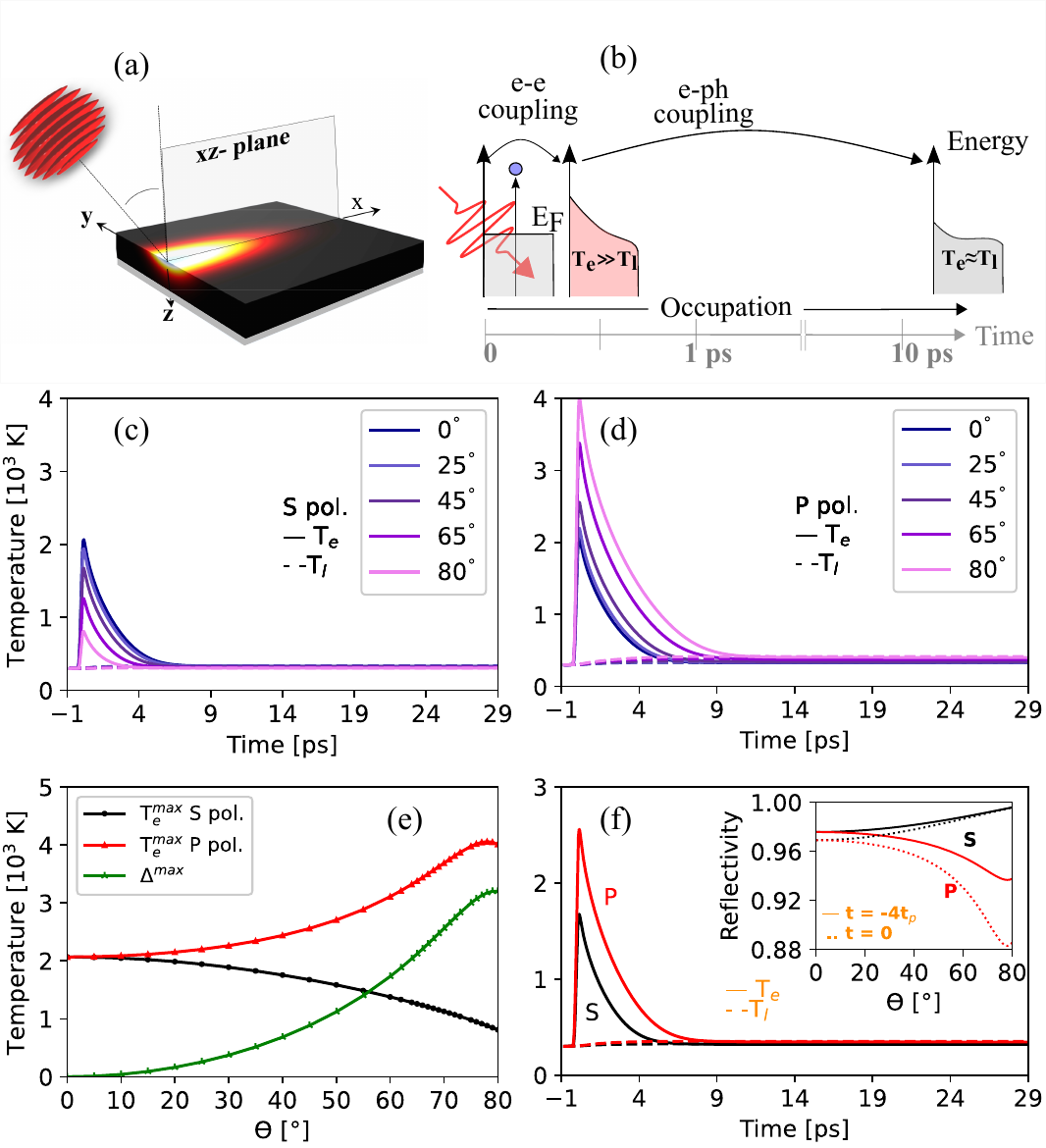}
\caption{(a) Schematic of the 3D irradiation geometry for ultrashort laser pulse focusing on metal film. (b) The time scales of ultrafast thermalization dynamics of the electron and the lattice sub-systems represented by the temperatures $T_{e}$ and $T_{l}$, respectively, driven out of equilibrium by the laser pulse. (c-d) Temporal evolution of the $T_{e}$ (solid lines) and $T_{l}$ (dashed lines) at the center of the laser-irradiated area for S and P polarization at different angles of laser incidence ($\theta$). (e) Maximum electron temperature $T_{e}^{max}$ for S (in black) and P polarized pulses (in red) and their difference (in green) as a function of laser incidence angle. (f) Temporal evolution of $T_e$ (solid) and $T_l$ (dashed) for S (in black) and P (in red) polarization for $\theta$ = 45$\degree$. The inset in (f) shows the reflectivity for S (black) and P (red) polarization as a function of $\theta$. Solid lines represent $R_{s,p}(T_{e}=300~K)$ and dotted curves represent $R_{s,p}(T_{e}=T_{e}^{max})$. The reflectivity minima for P polarization are observed at 79$^{\degree}$ near the Brewster's angle. Peak laser intensity $I_0$ = $4\times10^{11}~Wcm^{-2}$, pulse duration $t_p$ ( intensity FWHM) = $200~ fs$, wavelength $\lambda=800~nm$ , and thickness $d=500~nm$.}
\vskip -0.5cm
\label{figure_1}
\end{figure*}

\section{Theoretical model: 3D TTM coupled with space-charge limited thermionic emission}
In an experimental scenario, ultrashort laser pulse with an appropriate polarization is focused onto the relevant material surface at a predefined angle of incidence. The laser pulse temporal and spatial profiles, the focusing geometry as well as the material properties \cite{Madas2019} play a significant role in the interaction. In addition, the temperature dependent material characteristics evolve during the interaction and need to be dynamically incorporated in the description of the process, in a self-consistent manner. Experimental observables from such thermally dominated interactions are usually lumped parameters, like transient reflectivity \cite{Block2019,Unikandanunni2022,Block2023,Bresson2020}, ablation threshold  \cite{Cheng2016,Lizunov2022,Qiu2023,Sakurai2021} or thermionic electron emission yield \cite{Riffe1993,Li2019}, etc, which are intricately interlinked with time dependent material characteristics. For a finite laser focal spot size both the thermal evolution as well as the temperature dependent material parameters become spatially dependent necessitating a complete 3D description of the interaction. Thus, in our case for modeling the ultrafast energy deposition dynamics on metal surface and the subsequent thermal processes, we use an appropriate time dependent 3D model. 

\par The irradiation configuration is depicted in Figure \ref{figure_1}(a). The $(x,y,z)$ Cartesian coordinate system is fixed in such a way that $xy$ plane represent the plane of the irradiated surface and $z$ is orthogonal to the surface [defined by the plane $(x,y,0)$]. The origin $(x,y,z)=(0,0,0)$ is set on the target surface that receives the peak laser fluence. The laser pulse is focused onto the target surface at an oblique angle of incidence $\theta$, with $xz$ as the plane of incidence. The polarization of laser is oriented either in the plane of incidence (P) or perpendicular to it (S).

\par Subjected to ultrashort, moderately intense laser excitation (Peak laser intensity $I_0$ = $10^{9}\sim10^{14}~Wcm^{-2}$  and duration $t_{p}\in[100,1000]~fs$), a metal surface exhibits different thermal responses for the electronic and the lattice sub-system: spanning over 3-4 order of magnitude in $t_{p}$. Each of these processes operates on distinct time scales [Figure \ref{figure_1}(b)], and directly affect the underlying energy redistribution dynamics. Upon irradiation, the electron sub-system thermalize due to electron-electron ($e-e$) interaction typically on a time scale of $\tau_{ee}<100~fs$. The TTM does not describe the nonequilibrium dynamics of electrons that ensues, when irradiated by a pulse with duration $t_{p}$ significantly less than $\tau_{ee}$. Under such conditions the electron energy distribution deviates momentarily from the equilibrium Fermi-Dirac \cite{Block2019,Budai2022,Riffe2023} distribution. The maximum electron temperature can reach upto many thousands of Kelvin, however, within  the  duration  of  a single ultrashort pulse, the change of lattice temperature is generally negligible. Eventually, within $\tau_{ee}$, the electrons reestablish the  Fermi-Dirac distribution, through electron relaxation dynamics. In our case, we keep the pulse duration $t_p$ (intensity full width at half maxima or FWHM) $= 200~ fs$. Thus, upon laser excitation, the electron sub-system undergoes rapid local thermalization, and absorbed energy goes into increasing the temperature $T_{e}$ of the electron sub-system, driven out of equilibrium from the phonon background (or the lattice sub-system with a temperature $T_{l}$). The subsequent spatio-temporal dynamics is described by the two temperatures corresponding to the electron ($e$) and the lattice ($l$) sub-systems, respectively mediated by the electron-phonon ($e-ph$) coupling as presented in the schematic of Figure \ref{figure_1}(b). Eventually, the equilibrium between the electron and the lattice sub-system is restored at an elevated common temperature, typically on the time scale of several $ps$. 

\par The temperatures $T_e$ (for electron sub-system), $T_l$ (for lattice sub-system), and $T_s$ (for glass substrate) are functions of three dimensional space at each time step, and their evolution is governed by set of coupled differential equations that account the coupling between the thermal sub-systems, thermal diffusion and the laser absorption. The evolution of $T_e$, $T_l$, and $T_s$ are described by the following coupled non-linear partial differential equations, respectively:

\begin{subequations}\label{TTMeqs}
\begin{align}\label{el_subsystem} 
{C_e}(T_e)\frac{\partial T_e(x,y,z,t)}{\partial t}& = \nabla \cdot \left({k}_{e}(T_e,T_l)\nabla {T}_{e}(x,y,z,t)\right)-\\ \nonumber &G\big({T}_{e}(x,y,z,t)-{T}_{l}(x,y,z,t)\big)+\\ \nonumber &Q(x,y,z,t)-S_{es}(T_e,T_s)~,
\end{align}

\begin{align}\label{lat_subsystem}
{C}_{l}&\frac{\partial {T}_{l}(x,y,z,t)}{\partial t} = \nabla \cdot \left({k}_{l}\nabla {T}_{e}(x,y,z,t)\right) +  \\ \nonumber &G\big({T}_{e}(x,y,z,t)-{T}_{l}(x,y,z,t)\big)-S_{ls}(T_l,T_s)~,
\end{align}

\begin{align}\label{substrate_subsystem}
{C}_{s}\frac{\partial {T}_{s}(x,y,z,t)}{\partial t} &= \nabla \cdot \left({k}_{s}\nabla {T}_{s}(x,y,z,t)\right) +\\ \nonumber  &S_{ls}(T_l,T_s) + S_{es}(T_e,T_s)~,
\end{align}
\end{subequations}

where $C_e$, $C_l$, and $C_s$ denote specific heat capacities, and $k_e$, $k_l$, $k_s$ denote thermal conductivity for electron, lattice, and glass substrate sub-systems, respectively. Parameter $G$ accounts for the coupling between the electron and lattice systems. $S_{es}$($S_{ls}$) is the energy exchange parameter at the metal-glass interface for the metal electrons(lattice). 
 
\par $Q$ is the laser power density representing the laser energy input into the electron sub-system. We model the incident laser field within this source term as Gaussian, both in space and time (following \cite{ZhongyangWang1997}), focused obliquely at an angle $\theta$ (measured with respect to the target normal) on the surface of the 3D volume. Laser field incident on the surface is transformed to the surface coordinate ($x,y$) and the field ($E(x,y,t)$) and intensity distribution on the target surface is obtained, at each time step. This yields the intensity $I(x,y,t) = \epsilon_0 c/2|E(x,y,t)|^2$, where $\epsilon_0$ is the vacuum permittivity, $c$ is the speed of light in vacuum. However, at oblique incidence, to conserve the deposited energy that is spread out over an area greater than that of normal incidence, the intensity is multiplied by a factor of $1/\cos\theta$. However, in this study, we are conserving the deposited intensity and the $1/\cos\theta$ factor is not necessary to include. Thus, the effective intensity \enquote*{$I_{\text{eff}}$} deposited on the sample can be written as $I_{\text{eff}}(x,y,t) = \epsilon_0 c/2 |E(x,y,t)|^2$. This also requires the incorporation of an appropriate model of laser absorption at the immediate vicinity of the metal surface including a temperature dependent reflectivity $R_{s,p}(x,y,t)$ (s and p represent light polarization perpendicular and parallel to the $xz$ plane respectively in Figure \ref{figure_1}(a)), which is described later. $Q$ exponentially decays as a function of $z$, according to physical laws governing penetration of the pulse into the target.
\begin{align}\label{key3}
Q\left(x,y,z,t\right)=&\frac{I_{abs}(x,y,t)}{(\delta + \delta_b)(1-\exp{(-d/(\delta+\delta_b))})}\\ \nonumber & \times\exp\bigg(-\frac{z}{(\delta + \delta_b)}\bigg),
\end{align}
where $I_{abs}$ is the absorbed laser intensity on the sample at a point $(x,y)$ and is given by $I_{\text{abs}}(x,y,t) = [1- R_{s,p}(x,y,t)] I_{\text{eff}}(x,y,t)$. 

\par Thickness of Au film is given by \textit{d}$=500~nm$, thickness of substrate $d_{s}=500~nm$, $\delta$ is the wavelength and temperature dependent optical penetration depth, given as $\delta = 1/\alpha$ \cite{Wang2013}, where $\alpha = (4\pi Im(n_2))/\lambda$ is the absorption coefficient, $Im(n_2)$ is the imaginary part of the material's refractive index \enquote*{$n_2$}, discussed later. $\lambda$ is the central wavelength of the laser. To account for the impact of the ballistic motion and diffusion of hot electrons, the electron ballistic range $\delta_b=105~nm$ \cite{Wellershoff1999} is incorporated in addition \cite{SiemensNM2009} to the optical penetration depth \enquote*{$\delta$}. In our calculations, the analytical expression of temperature dependent electron-lattice coupling factor \textit{G} is given as \cite{Chen2005}  $G(T_e,T_l) = G_{RT}\Big[A_e/B_l(T_e+T_l)+1\Big]$, where $G_{RT} = 2.2\times10^{16}~Wcm^{-2}$ is the electron-lattice coupling factor at room temperature. $A_e = 1.2 \times 10^7~K^{-2} s^{-1}$  and $B_e = 1.23 \times 10^{11}~K^{-1} s^{-1}$  are the material constants. Constant values of $C_l$ = $2.45\times 10^{6}~Jm^{-3}K^{-1}$  \cite{mceuen2005introduction}, $C_s$ = $1.848\times 10^{6}~Jm^{-3}K^{-1}$  \cite{Block2019}, $k_s = 0.8~ Wm^{-1}K^{-1}$ \cite{SousaCastillo2017} are considered in this study. Temperature dependent parameters, such as $C_e(T_e)$ =$\gamma T_e$, $\gamma = 71~Jm^{-3}K^{-2}$,  $k_e(T_e,T_l)$ = $k_0(T_e/T_l)$, with $k_0 = 317~Wm^{-1}K^{-1}$   \cite{kittel1996}, and $k_l$ = $k_{eq}\times0.01~Wm^{-1}K^{-1}$  \cite{Zhang2008,klemens1986thermal}, where $k_{eq} = 320.973 - 0.0111~ T_l - 2.747\times 10^{-5} ~T_l^2 - 4.048 \times 10^{-9}~ T_l^3$ is the thermal conductivity of gold at equilibrium  \cite{touloukian1970thermal} are considered. $S_{es/ls} = G_{es/ls}(T_{e/l}-T_s)$ denotes the boundary interface heat transfer between the metal electrons/lattice and the glass substrate. Here, $G_{es} = (96.12 + 0.189~T_e)~MWm^{-2}K^{-1}$  and $G_{ls} = 141.5~MWm^{-2}K^{-1}$  \cite{Lombard2014} represent the thermal boundary conductances \cite{Giri2019,Sandell2020} for the electron-substrate and lattice-substrate systems, respectively.

\subsection{Initial and boundary conditions and thermionic emission rate}	
In order to solve the set of coupled non-linear partial differential equations given in \cref{TTMeqs}, the initial conditions are taken to be that the electronic, lattice, and substrate temperatures are at room temperature prior to the interaction, i.e. $T_e(x,y,z,-4t_p) = T_l(x,y,z,-4t_p) = T_s(x,y,z,-4t_p) = 300 ~K$, where $t_p$ is the intensity full width at half maximum (FWHM) duration of the laser pulse, and time $t = -4t_p$ is defined as the starting time of the simulation where the laser-sample interaction is yet to begin. The peak of the laser pulse arrives at the metal surface at time $t = 0$. At each time step, the thermionic emission is considered as the surface electron heat loss boundary condition on the irradiated surface of the sample \cite{Du2011,Balasubramni2009,Du2012}, and given as $k_e ~\partial _z {T}_{e}\Big|_{z=0} = -\big(eE_f+e\phi\big)\dot{N}_{sc}\Big|_{z=0}$, where \textit{z} = 0 is the irradiated surface, \textit{e} is the charge of the electron, $E_f$ is the Fermi energy, $\phi$ is the work function (work function of gold is $5.17~ eV$ \cite{Kim2017}) and $\dot{N}_{sc}$ is the space-charge limited thermionic emission rate per unit area (number/$m^2s$). The boundary conditions on the un-irradiated sides are implemented by assuming these surfaces to be thermally insulated from the ambient environment. At each time step, \cref{TTMeqs} is propagated with the above mentioned conditions and $\dot{N}_{sc}$ is calculated self-consistently from MRD. 

\par In the standard Richardson-Dushman equation \cite{Modinos1984}, the thermionic emission rate is given by $\dot{N}_1 = (A_0/e) T_e^2 \exp{\big(-e \phi/k_B T_e\big)}$, where $A_0 = 1.2 \times 10^6~A m^{-2} K^{-2}$  and $k_B$ is the Boltzmann constant. This expression assumes that, $E_f$ ($5.53~ eV$ for gold \cite{Billings1972}) is approximately equal to the free-electron chemical potential ($\mu$), which holds true when $k_BT_e$ is less than $1~eV$. Furthermore,
the space charge potential due to the electron emission is neglected. As the thermal evolution initiates the electron emission process, the emitted electrons leave behind positive residual charge on the metal surface, affecting further evolution of the emission current. Under femtosecond laser irradiation conditions, such effects, termed space charge effects can become significant \cite{Tao2017}. 
Understanding the strength of space charge effects is central to the correct estimation and interpretation of the laser-induced electron emission. Although, time-integrated total thermionic emission yield has been measured in many past experiments to make statements on strong space-charge suppression, the dynamic variation of the focal spot Gaussian contour and resulting ultrafast temporal variation of emission current has not been captured in previous studies. With growing possibilities to resolve and detect electron emissions at ultrafast time scale through current state-of-the-art experiments \cite{Caruso2019,Miller2014,Frster2016,Berger2012,Tan2017}, appropriate estimation of the space-charge potential is critical to predict the temporal and spatial properties of the emitted electrons.  

\par At higher temperatures, $\mu$ deviates from its value ($E_f$) at absolute zero. Thus, a suitable form of temperature dependent $\mu$ is essential. Furthermore, during thermionic emission, the emitted electrons form a thin disk of negative space charge cloud parallel to the metal surface \cite{Riffe1993}. The contribution of the space charge in the form of an effective potential $\phi_{sc}$ should also be taken into account in calculating the effective thermionic current density \enquote*{$\dot{N}_{sc}$}. Hence, we use a modified Richardson-Dushman equation, including the temperature-dependent chemical potential $\mu (T_e)$ and space charge potential \cite{Riffe1993} due to the thin disk of electrons near the solid surface, for calculating $\dot{N}_{sc}$: 
\begin{equation}
    \dot{N}_{sc}\left(x,y,t\right) = C (k_B T_e)^2 \exp\bigg[-\frac{eE_f-\mu (T_e)+ e\phi + \phi_{sc}}{k_B T_e}\bigg], 
\label{eqn:modRD}
\end{equation}
where $C=4\pi m/h^3$, $m$ is the mass of electron, $h$ is the Planck constant, $\phi_{sc}$ is the effective space-charge potential. $T_{e}$ in the above equation represents the surface electron temperature $T_e\left(x,y,z=0,t\right)$.

\subsection{Temperature Dependent Index of Refraction}
As mentioned previously, during the interaction, $I_{abs}$ in \cref{key3} is calculated from the reflectivity $R_{\mathrm{s,p}}$,		
which is given by the well-known Fresnel equations for S ($R_{s}$), and P-polarized cases ($R_{p}$) \cite{born1980basic}:
	\begin{subequations}
		\begin{align}
		R_{\mathrm{s}}(x,y,t)&=\left|\frac{n_{1} \cos \theta-n_{2} \sqrt{1-\left(\frac{n_{1}}{n_{2}} \sin \theta\right)^{2}}}{n_{1} \cos \theta+n_{2} \sqrt{1-\left(\frac{n_{1}}{n_{2}} \sin \theta\right)^{2}}}\right|^{2}, \\
		R_{\mathrm{p}}(x,y,t)&=\left|\frac{n_{1} \sqrt{1-\left(\frac{n_{1}}{n_{2}} \sin \theta\right)^{2}}-n_{2} \cos \theta}{n_{1} \sqrt{1-\left(\frac{n_{1}}{n_{2}} \sin \theta\right)^{2}}+n_{2} \cos \theta}\right|^{2},
		\end{align}
		\label{fresnel}
	\end{subequations}
where $n_1$ is the refractive index of ambient medium, assumed to be unity (corresponding to vacuum), and $n_2$ is the target's refractive index ($n_2 = n+ik = \sqrt{\epsilon_{DCP} (\omega)}$, where $\epsilon_{DCP}$ is the complex dielectric constant for gold; \textit{n} and \textit{k} are the real and imaginary parts of the refractive index, respectively).  
 
\par The refractive index depends on the wavelength of the incident laser pulse. However, a material's index of refraction may also change during periods of extreme temperature variations that occur during the interaction. For metals the widely used formula for calculating permittivity is derived from Drude-Lorentz model. It has been shown that the accuracy of the dielectric function can significantly increase, specially for noble metals like gold and silver, by utilizing the Drude-critical points (DCP) model \cite{Vial2008}. Hence, for our calculations, we use a model of gold's electric permittivity $\epsilon_{DCP}$ proposed in the existing literature \cite{Etchegoin2006,Etchegoin2007,Vial2008}. In addition, lattice and electron temperatures would have influence on the permittivity.  In order to take these effects into account and to  create a fully temperature-dependent model of index of refraction we have combined the temperature dependent electron relaxation times \cite{Block2019} with the Drude-critical points model. The model combines the Drude (intraband) contributions to permittivity with two \enquote{critical-points} modeling interband transitions \cite{Vial2008}:
 \begin{align}\label{eqn:vials-model}
	\epsilon_{\mathrm{DCP}}(\omega)  =\epsilon_{\infty}-\frac{\omega_{P}^{2}}{\omega^{2}+i \gamma \omega} +\sum_{p=1}^{2} &A_{p} \Omega_{p}\Bigg(\frac{e^{i \phi_{p}}}{\Omega_{p}-\omega-i \Gamma_{p}}+\\ \nonumber &\frac{e^{-i \phi_{p}}}{\Omega_{p}+\omega+i \Gamma_{p}}\Bigg),
\end{align}
where the first and second terms on the right hand side of \cref{eqn:vials-model} are the standard contribution of a Drude model \cite{bornprinciples} with a high frequency limit dielectric constant $\epsilon_{\infty}$, $\omega_P$ is the plasma frequency, $\omega$ is the driving laser frequency, and $\gamma$ is the damping term. The third term corresponds to the two interband transitions for which we use two critical points transition model explained in Leng et al. \cite{Leng1998}. The symbols $\phi_p$ is the oscillator phase, $\Omega_{P}$ is the oscillator energy, $A_p$ is the oscillator strength, and $\Gamma_p$, is the oscillator damping, given as $\Gamma_p$ = A$_{\Gamma p}$T$_e^2$ + B$_{\Gamma p}$T$_l$ + $\gamma_p$. The permittivity in \cref{eqn:vials-model} is given as a function of frequency of the incident light. Vial and co-workers \cite{Vial2008} obtained the constants $A_p$, $\Omega_p$, $\Gamma_p$, $\gamma$, $\phi_{p}$, and $\omega_{P}$ by fitting the model to experimental, room-temperature data from Johnson and Christy \cite{Johnson1972}. In previous works \cite{Etchegoin2006,Etchegoin2007}, the fit with all the free parameters produced $\phi_1 \sim \phi_2 \sim -\pi/4$. We fixed the two phases to be same and equal to $-\pi/4$ to have two parameters less and hence improve the convergence of rest of the parameters. 

\par However, this model is not temperature dependent. Hence, we made modifications in the model by considering $\gamma$ to be similar to the work by Block et al. \cite{Block2019} and $\Gamma_p$ to be temperature dependent,  and fitted our modified model to the data obtained by Johnson and Christy \cite{Johnson1972}. Considering just the Drude model term, the plasma frequency $\omega_p$ and the electron relaxation rate $\gamma$ actually vary with temperature. The plasma frequency is given by \cite{Block2019} $\omega_{P} = \sqrt{ (e^2 n_e(T_0))/(\varepsilon_0  m_{eff}(1 + \beta\Delta T_l)) }$, where $T_0$ is the ambient temperature, $n_e (T_0) = 5.9\times10^{28}~ m^{-3}$ is the Au free electron density \cite{ashcroft1976introduction}, $\varepsilon_0$ is the dielectric permittivity of the vacuum, $m_{eff}$  is the effective mass of electron which is estimated to be equivalent to 1.094 times the mass of electron $m_e$, and $\beta$ is the thermal expansion coefficient of Au, which is 4.23 $\times 10^{-5}~K^{-1}$ \cite{ashcroft1976introduction}. However, $\gamma$ is not well modeled by a constant value for systems with highly variable temperatures. In fact, this electron collision rate term can be separated into electron-electron collision rate $\gamma_{e-e}$ and electron-phonon collision rate $\gamma_{e-ph}$, where the rate of both types of collisions is $\gamma = \gamma_{\text{e-ph}} + \gamma_{\text{e-e}}$, where $\gamma_{\text{e-e}}$ = AT$_e^2$ and $\gamma_{\text{e-ph}}$ = BT$_l$ \cite{Block2019,Smith2001}. The analysis by Fisher et al. \cite{Fisher2001} provides an analytical estimation of both the terms. The parameters in \cref{eqn:vials-model} from the fit are displayed in Table \ref{Table1}.
\begin{table}[th!]
\small
\newcolumntype{C}{>{\centering\arraybackslash}c}
\centering
\begin{adjustbox}{max width=1\columnwidth}{\begin{tabular}{|c|c|}\hline
\rowcolor{BrightGray} \textbf{Parameter} & \textbf{Value} \tabularnewline \hline
\rowcolor{green!5} A [$K^{-2}s^{-1}$] & 1.2 $\times 10^{7}$  \tabularnewline \hline
\rowcolor{white}  B [$K^{-1}s^{-1}$] & 4.428703071 $\times 10^{11}$  \tabularnewline \hline

\rowcolor{green!5} $\Omega_1$ [$rad~ s^{-1}$] & 4.01772608$\times$10$^{15}$  \tabularnewline \hline
\rowcolor{white} A$_{\Gamma 1}$ [$K^{-2}s^{-1}$] & 1.2 $\times 10^{7}$  \tabularnewline \hline
\rowcolor{green!5}  B$_{\Gamma 1}$ [$K^{-1}s^{-1}$] & 1.14681587030 $\times 10^{11}$  \tabularnewline \hline
\rowcolor{white} $\gamma_1$ [$rad ~s^{-1}$] & 7.9 $\times$ 10$^{14}$  \tabularnewline \hline
\rowcolor{green!5}  $\phi_1$  & $-\pi/4$ \cite{Etchegoin2006,Etchegoin2007}  \tabularnewline \hline

\rowcolor{white} A$_{\Gamma 2}$ [$K^{-2}s^{-1}$]  & 1.2 $\times 10^{7}$  \tabularnewline \hline
\rowcolor{green!5}  B$_{\Gamma 2}$ [$K^{-1}s^{-1}$] & 6.094240955631399 $\times 10^{11}$  \tabularnewline \hline
\rowcolor{white} $\gamma_2$ [$rad~ s^{-1}$] & 1.9 $\times$ 10$^{15}$  \tabularnewline \hline
\rowcolor{green!5}  $\phi_2$  & $-\pi/4$  \cite{Etchegoin2006,Etchegoin2007} \tabularnewline \hline
\rowcolor{white} $\Omega_2$ [$rad~ s^{-1}$] & 5.56883141$\times$10$^{15}$  \tabularnewline \hline

\rowcolor{green!5}  $A_1$ & 0.917783355  \tabularnewline \hline
\rowcolor{white} $A_2$ & 1.52288304  \tabularnewline \hline
\rowcolor{green!5} $\epsilon_{\infty}$ & 1.197\tabularnewline \hline
\end{tabular}}
\end{adjustbox}
\caption{Parameters in \cref{eqn:vials-model} calculated by fitting to the data obtained by Johnson and Christy \cite{Johnson1972}. }
\label{Table1}
\end{table}
\par Thus, given $T_e(x,y,t)$ and $T_l(x,y,t)$ at the plane of interface, as well as the parameters of the incident laser, we can evaluate reflectivity $R_{s,p}(x,y,t)$. The calculations can be generalized and extended to any arbitrary orientation of the laser polarization, by decomposing the surface laser electric fields into orthogonal S and P components and combining the reflection coefficients of each polarization condition individually.

\section{Results and discussion}
\subsection{Dependence of electron and lattice temperature evolution on light polarization and angle of incidence}
\begin{figure}[th!]
\centering
\includegraphics[width=.47\textwidth]{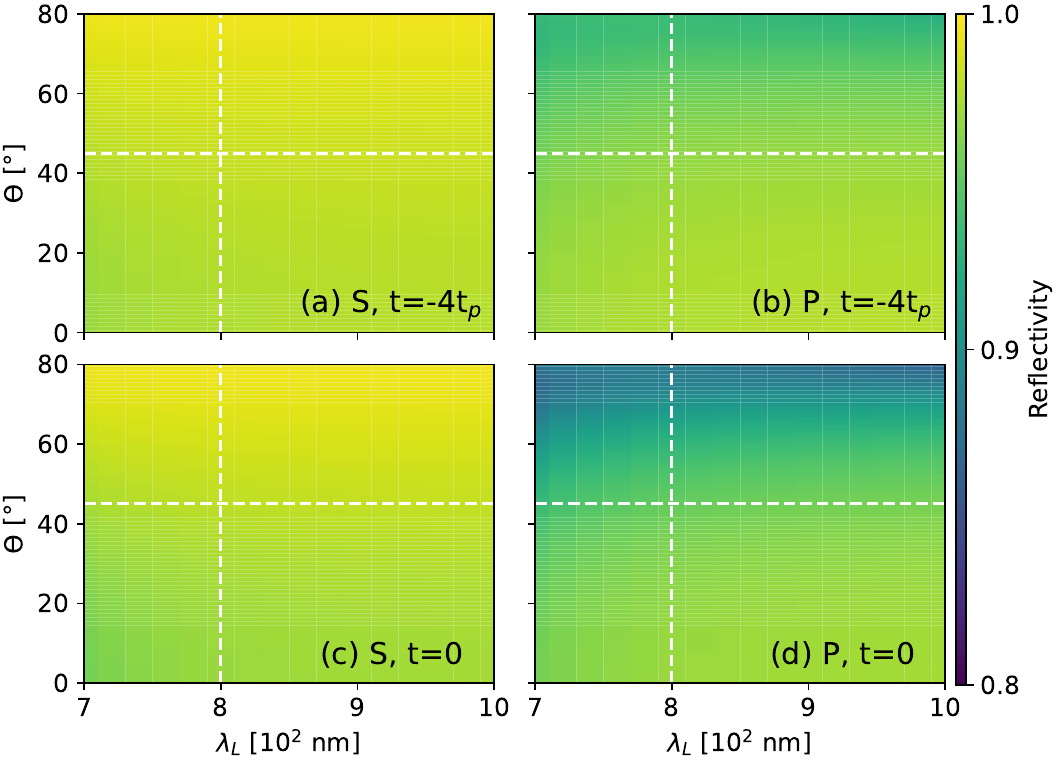}
\caption{(a,b) Reflectivity as a function of laser wavelength and angle of incidence for S and P polarized laser at $t = -4t_p$, $R_{s,p}(T_e=300~K)$  and (c,d) at $t = 0$, $R_{s,p}(T_e=T_e^{max})$, at the peak of laser intensity respectively.}
\vskip -0.35cm
\label{figure_2}
\end{figure}
In order to control and optimize the thermal evolution and to understand its dependence on the laser polarization and angle of incidence, we undertake a series of numerical simulations. Figure \ref{figure_1}(c), and \ref{figure_1}(d) show the temporal evolution of surface electron and lattice temperatures at the centre of laser-irradiated area for S and P-polarized laser respectively, as a function of varying laser incidence angle $\theta$. In all our simulations, laser pulse with an intensity full width at half maximum duration (FWHM) of $t_{p}=200~fs$, central wavelength of $800 ~nm$, peak intensity of $4\times10^{11}~Wcm^{-2}$ , beam waist size of $9 ~\mu m$ at FWHM of the peak intensity are considered, unless specified otherwise. 

\par At a given angle of incidence, for both the polarization, the electron temperatures peak almost at $t_{p}$ after the peak of the laser pulse has interacted with the target. The lattice temperature however does not increase significantly within the laser pulse duration, but slowly increases at the cost of cooling down the electrons over several $ps$, ultimately reaching thermal equilibrium with them. For the S-polarized case, the peak electron temperature $T_{e}^{max}$ decreases with increasing angle of incidence ($\theta\in [0\degree,80\degree]$) (Figure \ref{figure_1}(c)), whereas, for P-polarization $T_{e}^{max}$ increases under the same conditions (Figure \ref{figure_1}(d)). Thus, at lower angles of incidence, the results become less sensitive to the laser polarization, converging at $\theta=0\degree$, as expected since at normal incidence there is no difference between S and P cases, both being parallel to the target surface. The lattice peak temperatures ($T_l^{max}$) show the same dependency on angle of incidence. We also note that the electron-lattice (e-l) thermalization duration (duration required for electrons and lattice sub-systems to reach thermal equilibrium) increases as $T_e^{max}$ increases. Thus, as seen in Figure \ref{figure_1}(c) and \ref{figure_1}(d), the increase in the value of laser incidence angle reduces the e-l thermalization duration in S-polarized case. On the other hand, for P polarized pulse case, e-l thermalization duration increases as $\theta$ increases from 0$\degree$ to 80$\degree$. Figure \ref{figure_1}(e) shows $T_e^{max}$ values for S (in black) and P-polarized (in red) cases as a function of laser incidence angle. We evaluated the difference \enquote*{$\Delta^{max}$} (in green) between maximum value of electron temperature in S and P-polarized cases, and found out that the peak $\Delta^{max}$ occurs when $\theta=79\degree$. 

\par To highlight the effect of laser polarization and to choose a laser incidence angle that is experimentally feasible, the rest of the results, unless stated otherwise, are calculated for $\theta=45\degree$. In Figure \ref{figure_1}(f), we show the temporal evolution of surface electron and lattice temperature at the center of the laser-irradiated region for both S and P-polarized cases. For S polarized case, $T_e^{max}$ of $1674~ K$ is attained at $0.14~ ps$ and a $T_l^{max}$ of $324~ K$ is reached. On the other hand, in P-polarized case, $T_e^{max}$ is $2557~K$ attained at $0.15~ ps$ and $T_l^{max}$ is $352~ K$. Correspondingly, the e-l thermalization duration for S polarized case is around $6~ ps$ which is lower when compared with $8~ ps$ in the case of P-polarization. 

\par The polarization sensitivity of $T_{e}^{max}$ along with its dependence on $\theta$ can be interpreted in terms of laser energy absorption into the gold-glass hetero-structure. The polarization dependent reflectivity from gold can be obtained from \cref{eqn:vials-model} and \cref{fresnel}, which enters \cref{TTMeqs} through the source term in \cref{key3}. As is evident, a reduced reflectivity leads to more energy transfer from the laser pulse to the electron sub-system within the skin depth of the gold film, leading to an elevated $T_{e}^{max}$. Inset in Figure \ref{figure_1}(f) shows the surface reflectivity for both S (black dotted curve) and P-polarized (red dotted curve) cases as a function of various laser incidence angles obtained at the peak of the laser intensity, i.e., at $t=0$. To emphasize the enhanced effect of temperature dependent reflectivity we also plot room temperature reflectivity, at $t=-4t_p$ in the same plot (solid curves: red for P and black for S). For P-polarization, in both the temperature dependent cases, the minima of reflectivity are observed at  $\theta$ = 79$\degree$ near the Brewster's angle. As our results show, the reflectivity is comparatively lower in P-polarized case, thus depositing more laser energy to the sample. Therefore, for all the $\theta$ values, except for $\theta=0\degree$ (where S and P-polarized laser are same), surface electron temperature in the case of P-polarization is higher than S-polarized case. In addition, self-consistently incorporating the temperature dependence in reflectivity results in higher electron temperatures.
 
\par Hence, in all the numerical simulations, we have considered temperature dependent reflectivity of the material in order to calculate the thermal response of it under the influence of an ultrashort laser pulse. In Figure \ref{figure_2}, we have highlighted the variation of reflectivity of gold for different values of laser incidence angle as a function of varying laser wavelength. Reflectivity calculated at room temperature ($300~ K$, i.e. at $t=-4t_p$) for S and P polarized laser is shown in Figure \ref{figure_2}(a) and (b), respectively. Whereas, Figure \ref{figure_2}(c) and (d) for S and P polarized laser, respectively, show results at a time $t=0$ (i.e. at the peak of the laser intensity). It is evident from this figure that reflectivity is a function of both $T_e$ and $T_l$ (both are dependent on space and time). Thus, it is unrealistic to consider a constant reflectivity value in determining the spatio-temporal thermal response of a material. In Figure \ref{figure_2}, the laser wavelength and the incidence angle considered in the rest of the simulations is highlighted with vertical and horizontal white dashed line, respectively. We also note here that, in our case, for the $200~fs$ pulse centered at $800~nm$, the gold surface reflectivity is almost flat  within the bandwidth of the pulse and does not show signatures of interband transition \cite{Kolwas2020}. Thus, without any loss of generality, we can use the temperature dependent reflectivity for $800~nm$ in all our calculations.

\begin{figure*}[th!]
\centering
\vskip -0.5cm
\includegraphics[width=.8\textwidth]{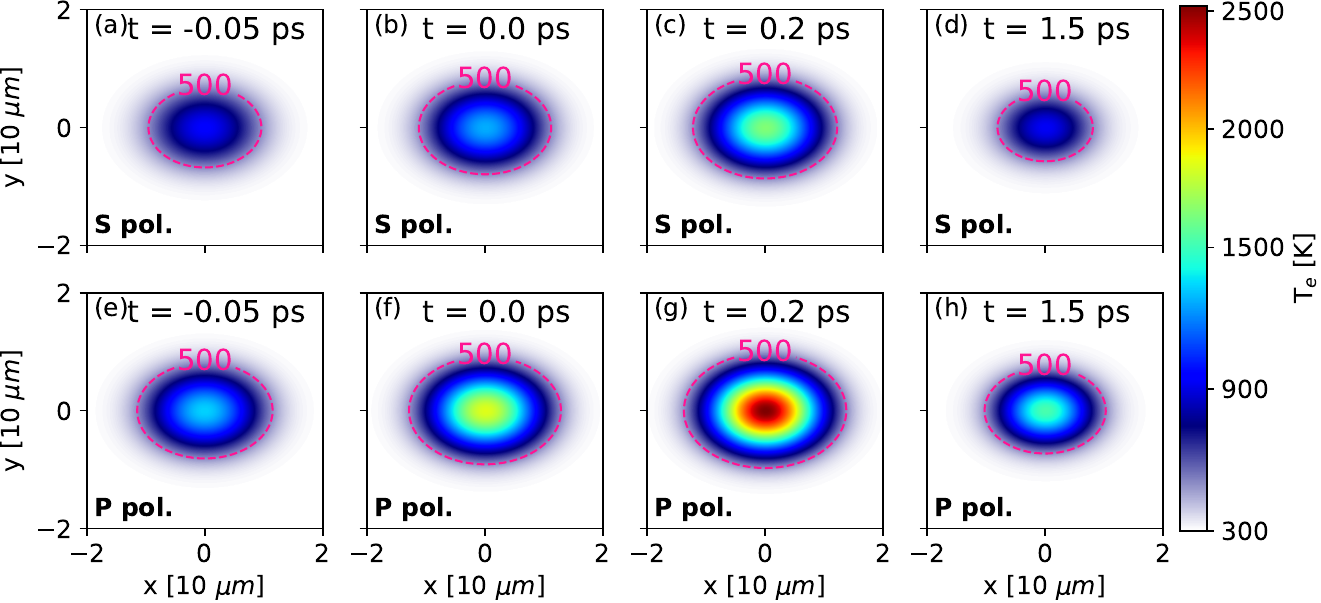}
\caption{Temporal evolution of surface $T_{e}$ for S-polarized (a-d) and P-polarized (e-h) laser pulse incident at an angle of 45${\degree}$ on to the sample. The pink iso-temperature line indicates $T_{e}=500~K$. The temperature profile is asymmetric along \textit{x} and \textit{y} axes, due to the oblique angle of incidence. The plotted colormaps show a sharp rise in electron temperature after laser energy deposition.}
\vskip -0.45cm
\label{figure_3}
\end{figure*}
\subsection{Spatio-temporal evolution of the surface electron temperature}
We calculated the spatial distribution of surface electron temperatures for both S and P-polarized cases at different time intervals just before and after the peak of the laser intensity (peak at $t=0~ps$) (see Figure \ref{figure_3}). The top (bottom) row in Figure \ref{figure_3} is obtained with S (P) polarized laser pulse. The difference in the electron temperature for S and P polarized case is evident in all the panels. We note here the asymmetry in the hot spot along the \textit{x} and \textit{y} direction. The elongation of the hot spot along the \emph{x}-direction is a signature of an elongated energy deposition profile due to the oblique incidence. In order to emphasize this aspect we have marked the $T_{e}(x,y,z=0)=500~K$ isocontour in red on each of the surface electron temperature profiles in Figure \ref{figure_3}. Beyond $t=0$ (at peak laser intensity), $T_e$ continues to rise until $t=0.15~ ps$, then starts to decrease. This behavior is consistent with the result shown in Figure \ref{figure_1}(f), where the rise of $T_e$ is sharp, while its fall beyond $0.15 ~ps$ is comparatively slower. Prominent difference of $T_e^{max}$ for S and P polarized cases is visible from the temperature profile at the focal spots in Figure \ref{figure_3}(c) and (g), as also discussed previously in the context of Figure \ref{figure_1}. The reason is attributed to different reflectivity conditions for S and P polarized laser, all across the fluence profile on the irradiated gold surface. The evolution of the surface temperature of laser induced hot spot presented in Figure \ref{figure_3}(a-h) contains vital information for time resolved pump-probe experiments like in 2D mapping of the transient reflectivity \cite{Pan2020}. Such experiments on a metal nano-film coated surface can be conducted in state of the art beamlines \cite{Mondal2018} in facilities like ELI-ALPS \cite{Charalambidis2017}.

\par In order to investigate the evolution of the lateral size of the laser generated hot spot, we take lineouts along \textit{x} and \textit{y} axes on the surface electron temperature profile, at different time instants. Corresponding \textit{x} and \textit{y} lineouts through the center of the hot spots of the $T_e$ profiles from Figure \ref{figure_3} are presented in Figure \ref{figure_4}. Figure \ref{figure_4}(a) and (c) are the $T_e$ evolution variation along \textit{x} with $y=0$ and variation along \textit{y} with $x=0$, respectively, for the S-polarized pulse case. Whereas, Figure \ref{figure_4}(e) and (g) represent the same for the P-polarized case. At each time $t$, we then evaluate the half width half maximum (HWHM) lengths of the electron temperature (Gaussian-like) profile along \textit{x} direction ($R_1$) and along \textit{y} direction ($R_2$). Calculated values of both $R_1$ and $R_2$ at each time $t$ are shown in the Figure \ref{figure_4}(b) and (d) for S-polarized case and in Figure \ref{figure_4}(f) and (h) for P-polarized case. The blue dashed curves presented in Figure \ref{figure_4}(b), (d), (f) and (h) are the variations in $R_1$ and $R_2$ for the corresponding polarization scenarios. The colormap in Figure \ref{figure_4} represent the time scale and the colored circles presented in the panel on the right column are related to the corresponding colored curves on the left column. The evolution of the surface hot spot size shows the same behaviour for both the S and P polarization and it qualitatively follow the $T_{e}$ dynamics presented in Figure \ref{figure_1}(f). In each case, the hot spot size first increases rapidly in sub-ps time scale immediately after laser excitation to its maximum value and then slowly decay down over several \emph{ps} eventually saturating to a constant value once electron lattice thermalization is complete. We also note that the peak hot spot size reached in each case ($2R_{2}\sim 10-11~\mu m$) at the peak of the electron temperature, is little higher than the focal spot size (FWHM =$9~\mu m$), but $R_{2}$ reduces to $\sim 6-7~\mu m$ over several \emph{ps} time scale. Estimations based on such calculations is of paramount importance in both designing experiments for studying heat dynamics in plasmonics \cite{Cunha2020} and for applications in pump-probe thermoreflectance \cite{Poopakdee2022} configuration.

\par We would like to comment here that, under conditions where the electron temperature is comparable or higher than the Fermi temperature (Gold Fermi temperature is $T_F=6.42\times10^4~K$ \cite{ashcroft1976introduction}), the TTM needs to incorporate full quantum mechanical calculations \cite{Jiang2005}. In order to access dynamics over a much longer time-scale and for comprehensive details one would need to use computationally costly molecular dynamics simulations coupled to 3D TTM \cite{Ivanov2003,Wu2013,Arefev2022}. For more rigorous calculations enabling a parameter free description, TTM model can be replaced with \emph{ab-initio} time-dependent Boltzmann equations \cite{Caruso2022}. Nonetheless, under the conditions addressed in this investigation our formulation provides a reasonably valid description.

\subsection{Spatio temporal evolution of the thermionic current density}
We compute the thermionic emission current density due to the thermal evolution of the laser heated metal surface by using MRD (\cref{eqn:modRD}) self-consistently together with the TTM at each time step, with the initial and boundary conditions mentioned previously. As initial condition on $\dot{N}_{sc}$, we assume that there is no electron emission from the gold surface prior to laser excitation. In this work, we consider the thermionic current based on a modified Richardson-Dushman equation,  after incorporating the space-charge field created by the disc of electron cloud near the laser-induced focal spot on the metal surface. We also consider the contribution due to the temperature dependence of the chemical potential. Before delving further on the results, we discuss the nontrivial terms representing potentials $\phi_{sc}$ and $\mu (T_e)$ inside the exponential in MRD. 

\subsubsection{Dynamic Space charge effects and corresponding validity regime of the model}

\par Thermionic emission ensues, soon after laser-induced electron heating of the metal surface. As the ultrashort thermionic emission starts taking place, a thin charged disk of escaped electrons parallel to the metal surface forms, introducing the space charge barrier. Consequently, suitable consideration of laser induced space-charge effects is required. The influence of the space charge field is in general complex \cite{Zhou2005,Darr2020},  all the emitted electrons need not experience the same barrier and thus not always it can be represented as an effective potential lumped into $\phi_{sc}$. The most rigorous treatment to deal with space charge effects require N-body numerical simulations incorporating Coulomb force calculations and a solution of equations of motion with sophisticated numerical schemes \cite{Barnes1986,Hellmann2009,Hellmann2012}. However, for ultrafast emissions, which is the case for ~\emph{fs} laser irradiation, the emitted electrons are well localized in space and are emitted early in the interaction allowing simplifications \cite{Riffe1993,Oloff2014}.

\par If the laser pulse is short enough the emitted electrons form a thin disk with spatial width $\Delta x$ parallel to the gold surface. To investigate further, we consider a rectangular temperature pulse of duration $\tau$ with maximum electron temperature $T_{e}$ and focus our attention on the emitted electrons that are still above the surface. If $\tau$ is also the emission duration then the maximum width of the emitted electron disk is $\Delta x=(3k_{B}T_{e}/m)^{1/2}\tau$ (where $(3k_{B}T_{e}/m)^{1/2}$ is the root mean square speed) and the maximum lateral width can be estimated by $max(2R_{1},2R_{2})=2R_{1}$. Thus, we can define the temperature dependent ratio  to set a criteria on whether the emitted hot electrons can be approximated by a thin disk or not, given as:
\begin{align}\label{sc_validity}
\eta(t) &= \Delta x(T_e(t))/2R_{1,2}(T_e(t)) \nonumber \\ &=(3k_{B}T_{e}/m)^{1/2}\tau/2R_{1,2}(T_e(t))~.
\end{align}
Thus, for $\eta(t) \ll 1$, the electrons emitted by thermionic process is well represented by a thin disk. Assuming $\tau$ in gold to be about $\sim1~ ps$ \cite{Riffe1993,Bauer2005}, the validity of the disk model has been tested. Using one temperature for the emitted electrons, the space charge barrier potential is given as $\phi_{sc} \approx aN_{yield}e^2/R_1$, where $N_{yield}$ is the analytical expression for total yield. For such a rectangular temperature pulse of duration $\tau$, Eq. \ref{eqn:modRD} can be integrated analytically \cite{Riffe1993,Wang1994,Du2011,Balasubramni2009} to give, $N_{yield}=(k_BT_e)/(ae^2/R_1)\log\big[1+C\tau\pi R_2ae^2k_BT_e\exp{\big(-(eE_f-\mu(T_e)+e\phi)/(k_BT_e)\big)}\big]$, where $R_1~\text{and}~R_2$ are the lengths of the semi-major and semi-minor axes of the elliptical electron charge disk and $a$ is a constant that depends on the geometry of the escaping electron cloud. For an uniform thin disk $a$ = $16/(3\pi)$ = 1.7 \cite{Riffe1993}. In this case, we note here that for $aN_{yield}e^2/R_1\ll k_{B}T_{e}$ the space charge contribution would be negligible and the expression of $N_{yield}$ reduces to the standard Richardson-Dushman equation, which would then be a valid description. The space charge effects would start to influence the emission rate and hence the yield as soon as $aN_{yield}e^2/R_1\approx k_{B}T_{e}$. Nevertheless, in a real scenario with a non-rectangular pulse shape, one would need to numerically evaluate $N_{yield}$ by integrating MRD.
\begin{figure}[th!]
\centering
\includegraphics[width=0.47\textwidth]{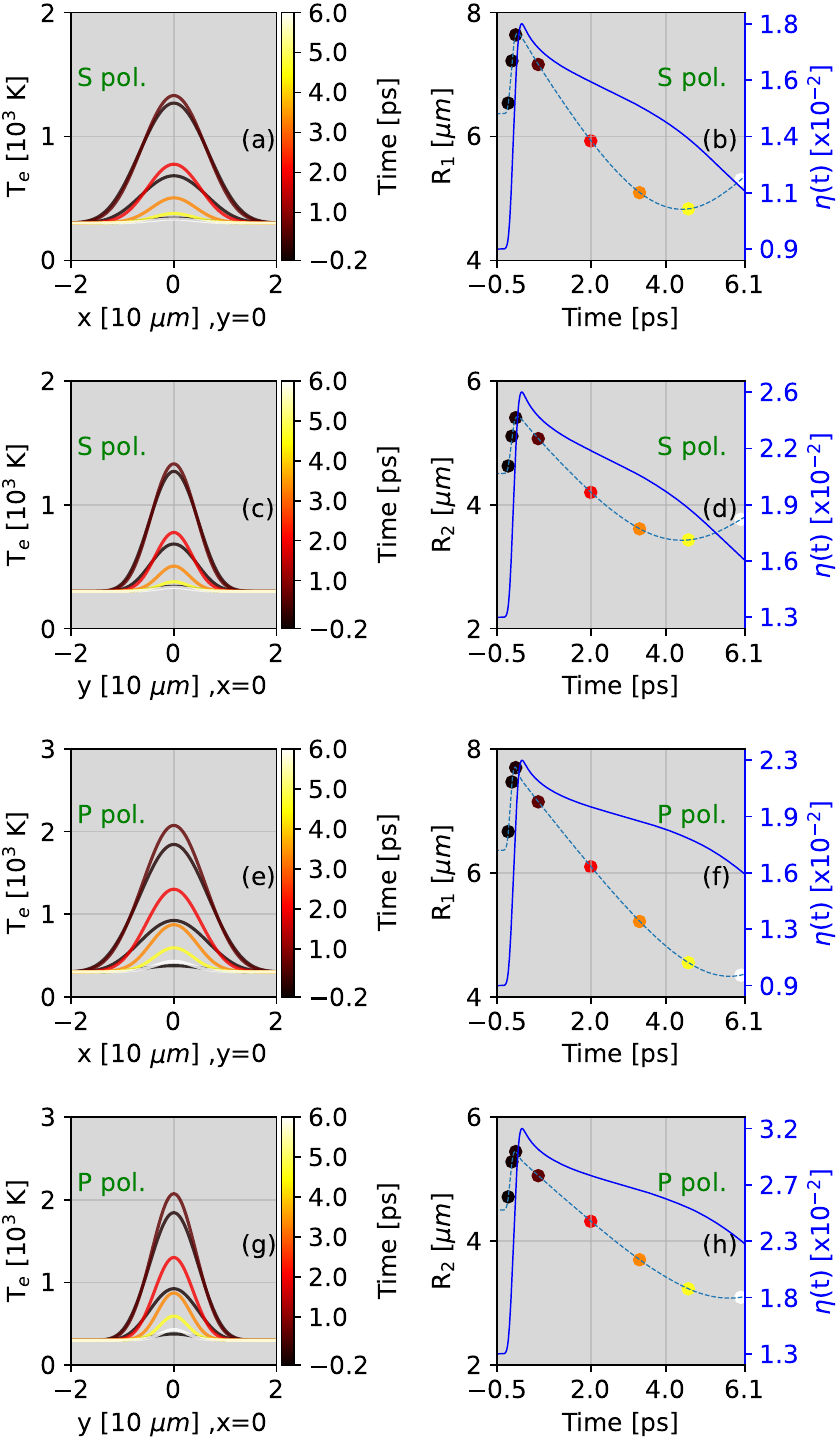}
\caption{The evolution of the \textit{x} and \textit{y} line-outs of surface temperature profiles at different instants of time for S (first and second row) and P-polarization (third and fourth row) incident on to the sample at $\theta=45\degree$. Corresponding HWHM of the $T_e$ distribution (Gaussian-like profile) along \textit{x} and \textit{y} directions, referred as $R_1$ and $R_2$, respectively are shown in the second column. The right axes in the second column (blue curve) represent $\eta(t)$ (as given in Eq. \ref{sc_validity}) for the validity regime of the space charge model. Input parameters are same as mentioned in Figure \ref{figure_3}.}
\vskip -0.5cm
\label{figure_4}
\end{figure}
\par Following the ansatz in \cite{Riffe1993}, under the thin disc condition, the model of space charge is valid if, $\eta(t) \ll 1$, where $2R_{1,2}$ is the lateral spatial extent depending on the direction of consideration. $2R_{1,2}(T_e(t))$ along \emph{x}-direction is $2R_1$ and along \textit{y}-direction is $2R_2$ as given in Figure \ref{figure_4}. Thus, parameter $\eta(t)$ provides the trend of temporal evolution of space charge and validity regime of the space charge model. The variations of $\eta(t)$ as a function of time, obtained from our simulations, for both S and P polarized laser fields are presented in blue solid lines in Figure \ref{figure_4}(b), (d), (f) and (h). In all the circumstances, $\eta(t)$ is two orders of magnitude smaller than 1, thereby validating the model for space charge incorporated thermionic emission to be extended to our case.

\par We extend the space charge model by incorporating spatial (and also temporal) dependence of $\phi_{sc}$, through the computed surface temperature profile. We calculate $\phi_{sc}$ based on the emitted electrons coming out of the surface of the sample. Thus, for our formulation, $\phi_{sc} = aN_{yield}e^2/R_1(t)$, thereby providing the dynamic variation of the space charge. The total number of thermionic electrons emitted per unit area \enquote*{$N_{sc}$} (number/$m^2$) is calculated numerically, and given as
\begin{equation}\label{Nsc}
    N_{sc} = \int_{t_i}^{t_f}{\dot{N}_{sc}}~ dt,
\end{equation}
where $t_i~\text{and}~t_f$ are the numerical simulation initial and final times, respectively. $N_{yield}$ is given as, $N_{yield}=\sum{N_{sc}}\delta x\delta y$, where $\delta x ~\text{and}~ \delta y$ are the grid sizes along the \textit{x} and \textit{y} directions, respectively.

\par Depending on the pulse profile, laser polarization, the angle of incidence and the focal spot size the thermal distribution on the target surface and hence the space charge accumulated over the metal surface will change. Thus, in our calculations of thermionic emission rate \enquote*{$\dot{N}_{sc}$}, we incorporate this contribution in a time dependent manner. Instead of taking a rectangular $T_{e}$ temporal profile, we take its profile self-consistently. 

\subsubsection{Temperature dependence of chemical potential}
\begin{figure}[bt!]
\centering
\includegraphics[width=0.48\textwidth]{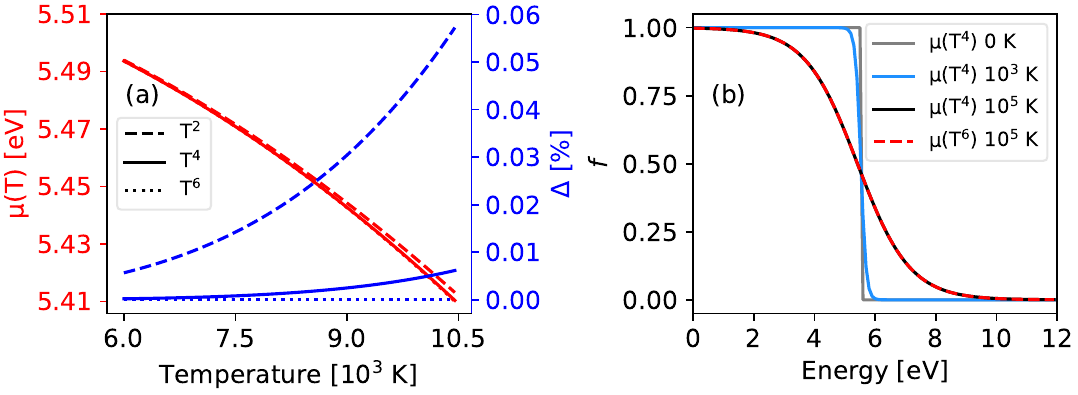}
\caption{(a) Temperature dependent variation of chemical potential $\mu(T)$ with the inclusion of terms up to the order of T$^2$ (dashed), T$^4$ (solid), and T$^6$ (dotted) in \cref{eqn:mu}. Right hand y axis (in blue color) is the percentage variation $\Delta$ of $\mu(T)$ for the corresponding cases. (b) Dependence of Fermi-Dirac distribution functions for gold on $\mu(\mathcal{O}(T^p))$, at different temperatures.}
\label{figure_5}
\end{figure}
According to the Fermi-Dirac distribution, at absolute zero temperature, the low-lying single particle states become occupied up to the Fermi energy, following the Pauli exclusion principle, while any states beyond $E_f$ remain unoccupied. 
As the temperature rises, the total internal energy of the system increases, consequently the electrons undergo excitation, thereby populating higher states beyond $E_f$.
Therefore, an increased number of low-lying single particle states become unoccupied, causing a decrease in the energy of the lowest lying states. Consequently, the chemical potential $\mu$ decreases. Thus, temperature dependent chemical potential should be included in calculations of $\dot{N}_1$. 

\par The Fermi-Dirac equilibrium distribution of electrons, is given by $f_i^N = 1/(\exp{\big((\epsilon_i - \mu)/k_BT\big)}+1) $, where $\mu$ is the chemical potential, $T=T_{e}$ is the temperature. $f_i^N$ is the mean number of electrons in one-electron level $i$. The total number of electrons $N$ is just the sum over all levels of the mean number in each level. Sommerfeld expansion can be applied to the integral forms to calculate the temperature dependence of the chemical potential \cite{ashcroft1976introduction}, which is given by,
\begin{align}\label{eqn:mu}
    \mu (T_e) = &eE_f\Bigg[ 1 - \frac{\pi^2}{12}\Bigg(\frac{k_BT_e}{eE_f}\Bigg)^2 \\\nonumber
    &- \frac{3\times 7\pi^4}{8\times 360}\Bigg(\frac{k_BT_e}{eE_f}\Bigg)^4 - \frac{31\times 105\pi^6}{15120\times 32}\Bigg(\frac{k_BT_e}{eE_f}\Bigg)^6 \Bigg]~.
\end{align}


\par The dependence of $\mu$ on higher order terms in $T$ (up to $T^6$), is evaluated from \cref{eqn:mu} and plotted in Figure \ref{figure_5}(a). The curves in red in Figure \ref{figure_5}(a) present the variations in $\mu$ with temperature when terms up to $\mathcal{O}(T^2)$ (dashed), $\mathcal{O}(T^4)$ (solid) and $\mathcal{O}(T^6)$ (dotted) are included. In order to observe the contributions of the higher order terms to the chemical potential, in Figure \ref{figure_5}(a) we also plot the percentage variations \enquote*{$\Delta[\%]$} of $\mu(T)$ (blue curves) using the expression, $\Delta = [\mu(\mathcal{O}(T^p))-\mu(\mathcal{O}(T^6))]/\mu(\mathcal{O}(T^6))$, where $p~=~2,~\text{or}~4,~\text{or}~6$. It is evident from the figure that, percentage variation $\Delta[\%]<0.01$ when high order terms up to $T^4$ is included for $T<10.5\times10^{3}~K$, which is the case in this study. Thus, for temperatures below $1 ~eV$, the higher order thermal contribution to the chemical potential is not significant. We further probe the influence of temperature dependent $\mu$ on the Fermi-Dirac distribution in Figure \ref{figure_5}(b). The resulting Fermi-Dirac distribution \enquote*{\textit{f}}, which depends on $\mu(T)$ is plotted as a function of energy in Figure \ref{figure_5}(b). Even though, by considering very high electron temperature, for instance, $T_e$ = 10$^5~K$  in $\mu(T_e)$, no significant change in Fermi-Dirac distribution is observed in both the cases $\mu(T^4)$ and $\mu(T^6)$ (as shown in Figure \ref{figure_5}(b)). Hence, higher order terms ($>\mathcal{O}(T^4$)) are neglected in \cref{eqn:mu} for our calculations of the thermionic emission, which we discuss in the next subsection. 
\begin{figure*}[th!] 
    \centering
    \includegraphics[width=.99\textwidth]{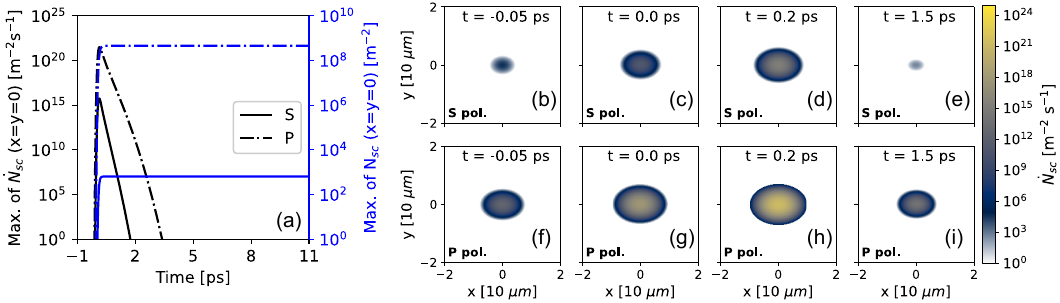}
    \caption{(a) Temporal evolution of thermionic emission rate \enquote*{$\dot{N}_{sc}$} (in black) which is maximum at the center of the irradiated surface ($x = y = 0$) obtained for both S (solid curve) and P (dash dot curve) polarized laser, calculated using \cref{eqn:modRD}. Subsequently, total number of thermionic electrons \enquote*{$N_{sc}$} (blue curves) emitted per unit area at the center of the irradiated surface ($x=y=0$) calculated using the \cref{Nsc} for S (solid curve) and P (dash dot curve) polarized laser. Snapshots of \enquote*{$\dot{N}_{sc}$} for S (b-e) and P polarized laser (f-i) at different time steps. Input parameters are same as mentioned in Figure \ref{figure_3}.}
   \label{figure_6}
\end{figure*}
\subsubsection{Evolution of the thermionic current and charge}

Under the same irradiation conditions as in Figure \ref{figure_1}(f) and in Figure \ref{figure_3}, now we investigate how the thermionic current density, flux and the emission profile on the laser spot on target behave. The temporal evolution of $\dot{N}_{sc}$ at the center of the laser-irradiated area is calculated using \cref{eqn:modRD} and presented in Figure \ref{figure_6}(a) (in black). Please note that the vertical axis here is plotted in log-scale (contrary to the linear scale used in Figure \ref{figure_1}(f) for the temperature). The results show that both for S (solid line) and P (dash-dotted) polarization the emission current density is ultrashort in nature (typical FWHM of the current pulse is $<~1 ~ps$). The trends demonstrate rapid rise of peak $\dot{N}_{sc}$ and higher value for P-polarized case, similar to $T_e$ evolution in Figure \ref{figure_1}(f). The thermionic emission tail is prolonged for P-polarized case, with a rise duration of $0.24 ~ps$ from $\dot{N}_{sc}$ value of $10^0~m^{-2} s^{-1}$ until the peak value and drop duration of $2.8 ~ps$ from the peak. For S-polarized case, the rise and drop durations are $0.2 ~ps$ and $1.6 ~ps$, respectively. The temporal evolution of $N_{sc}$ from the center of laser irradiated area obtained from Eq. \ref{Nsc} (blue curves in Figure \ref{figure_6}(a)) indicates prominent difference for S and P-polarized cases. After the thermionic current density pulse is over, the blue curves saturate, implying that the emission has been terminated causing no further change to the emitted number density.
The spatial distribution of $\dot{N}_{sc}$ is captured at the same time instances as considered in Figure \ref{figure_3} and presented in Figure \ref{figure_6}(b-i). Since $T_e^{max}$ is lower in S-polarized case compared to P-polarized case, $\dot{N}_{sc}$ is very low for S compared to P-polarized case, as is evident in all the panels of Figure \ref{figure_6}(b-i). A clear dependence of surface $T_e$ on $\dot{N}_{sc}$ is observed; higher the $T_e$, greater is the value of $\dot{N}_{sc}$. Due to the oblique laser incidence, the $\dot{N}_{sc}$ distribution is asymmetric along \textit{x} and \textit{y} directions.  
\begin{figure*}[th!] 
    \centering
    \includegraphics[width=.97\textwidth]{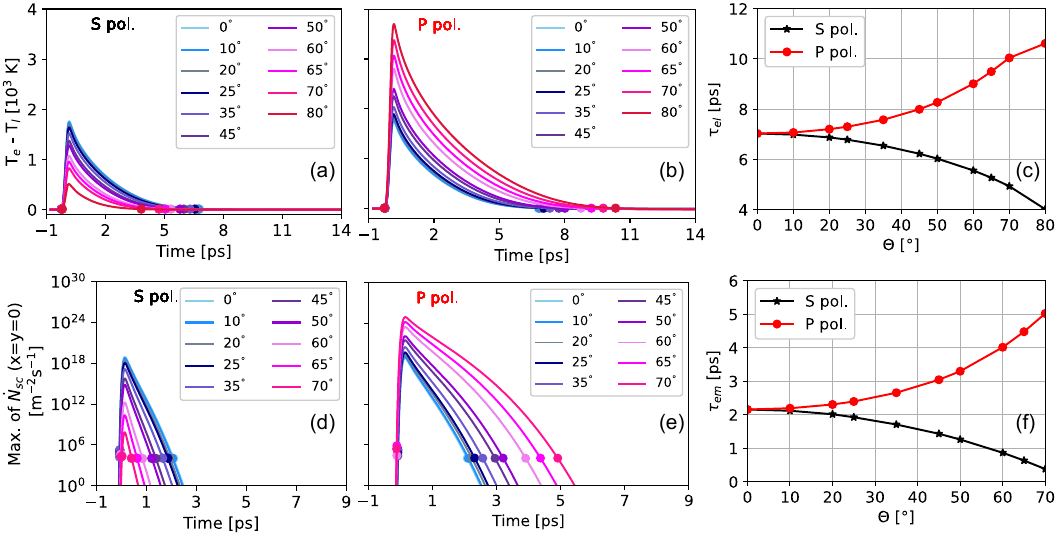}
    \caption{The difference between surface electron and lattice temperature at the center of the laser irradiated area for different laser incidence angles for S (a) and P (b) polarized laser pulse. The time between the two circle markers (located when $T_e-T_l=10 ~K$) for each laser incidence angle is considered to be the electron-lattice thermalization duration \enquote*{$\tau_{el}$}, which is represented in (c) for S (black curve) and P (red curve) polarized laser pulse. The thermionic emission rate at the center of the laser irradiated area for different laser incidence angles for S (d) and P (e) polarized laser pulse. The duration between the two circle markers for each laser incidence angle is considered to be the thermionic emission duration \enquote*{$\tau_{em}$}, which is represented in (f) for S (black curve) and P (red curve) polarized laser pulse.}
    \label{figure_7}
\end{figure*}
\par The dependence of S and P laser polarization on e-l thermalization duration \enquote*{$\tau_{el}$} and thermionic emission duration \enquote*{$\tau_{em}$} is calculated as a function of laser incidence angle and presented in Figure \ref{figure_7}. The difference in surface electron and lattice temperature ($T_e-T_l$) at the center of the laser irradiated area for S and P polarized laser is calculated as a function of laser incidence angle and presented in Figure \ref{figure_7}(a) and (b), respectively. $\tau_{el}$ is the duration required for the electrons to transfer their energy to lattice and attain thermal equilibrium. For each laser incidence angle, in order to obtain $\tau_{el}$, we selected two time instants where $T_e-T_l=10 ~K$ (indicated by circles in Figure \ref{figure_7}(a, b)) located before and after the peak of $T_e-T_l$. We evaluate the time difference between these two selected values and use this to represent $\tau_{el}$. The $\tau_{el}$, thus obtained, for S and P polarization cases are plotted as a function of $\theta$ in Figure \ref{figure_7}(c). With increasing angle of incidence, $\tau_{el}$ decreases for S polarization, whereas, in the case of P polarization it increases. This behavior could be attributed to the fact that as the electron temperature raises, more time is required for those hot electrons to transfer their energy to lattice. As it was shown earlier in Figure \ref{figure_1}(e), maximum of the surface electron temperature at the center of the laser irradiated area is higher in the case of P polarized laser when compared to S polarized case. Hence, $\tau_{el}$ is higher for P polarized case as shown in Figure \ref{figure_7}(c). The reason for this kind of behavior could be explained with the support of reflectivity result shown in the inset of Figure \ref{figure_1}(f). It is clear from the inset of Figure \ref{figure_1}(f) that the reflectivity in the case of P polarized laser decreases, which means that more amount of energy is absorbed by the sample, resulting in high $T_e$. The opposite is true in the case of S polarized pulse. Therefore, higher surface electron temperature is attained in the case of P polarized laser compared to S polarized case. Similar kind of trends seen in Figure \ref{figure_7}(a-c) are detected in thermionic emission duration $\tau_{em}$ shown in Figure \ref{figure_7}(d-f).

\par Due to low $T_e^{max}$ value in the case of S polarized laser compared with P polarization case (see Figure \ref{figure_1}(e)), the hot electrons takes less time to transfer their energy to lattice, subsequently  lowering their temperature, which results in less thermionic emission duration (Figure \ref{figure_7}(d)). On the other hand, due to higher $T_e^{max}$ in P polarized case, hot electrons require more time to transfer their energy to lattice, thereby due to the existence of hot electrons for longer period, thermionic emission takes place for more duration (Figure \ref{figure_7}(e)). Due to these effects, the thermionic emission duration ($\tau_{em}$) increases in P polarized case, and decreases in S polarized case as a function of laser incidence angle (Figure \ref{figure_7}(f)). $\tau_{em}$ is calculated by considering two time instants when peak of $\dot{N}_{sc}=10^4 ~m^{-2}s^{-1}$, as highlighted by filled circles. The duration between these two points represents the thermionic emission duration. 

\section{Summary and Conclusions}
Gold being a noble \cite{Balcerzak2021} transition metal has multifaceted generic applications ranging from flexible integrated electronics \cite{Takakuwa2021}, biomedicine \cite{Xu2021} to twistronics \cite{Heyl2022}. In addition, gold coated mirrors are ubiquitous in ultrafast laser optics. Ultrafast thermal management and electron emission metal nanofilms are of paramount importance in studies that extend all these applications. 
In the present study, we implemented and utilized space-time resolved numerical simulations to explore the polarization dependent ultrafast thermionic emission from a nanometric gold film coated glass target under oblique irradiation of a focused femtosecond laser pulse. We implemented an improved TTM that simulates volumetric heating incorporating dynamic optical and material properties, as well as the modified Richardson-Dushman equation to compute the thermionic emission profiles.

\par We examined the temporal evolution of electron and lattice temperatures with varying laser incidence angles. Our findings indicate that at near Brewster angle of incidence, laser polarization can be switched from S to P to control and increase the hot spot size, the surface electron temperature and the thermionic emission rate. Our analysis show that at all angles of incidence, the duration of thermionic current pulse shows a significant correlation with the intrinsic electron-lattice thermalization time of the sample. Since increase in electron phonon coupling strength leads to a reduction in electron-lattice thermalization time, our results indicate that by proper choice of metal coating, it would be possible to control the  pulse duration of the thermionic current, while maintaining the peak brightness. In view of the recent experimental capabilities to resolve and detect electron emissions at ultrafast time scale \cite{Caruso2019,Miller2014,Frster2016,Berger2012} such insights enable further lines of investigation. This also holds application potential for the development of high-brightness ultrafast electron imaging tools.

\par We also note that, thermophotovoltaic power generation has recently achieved significant gains in efficiency \cite{ LaPotin2022} under conditions of high emitter temperatures \cite{Mittapally2021} and the device sometimes use reflective metal-dielectric structure \cite{Fan2020}. Such approaches have significant implications for applications in energy harvesting. Moreover, thermionic emission based solar concentrators \cite{ Schwede2010, Xiao2017} and thermionic energy converters \cite{Campbell2021} also use metal based architectures. Nonetheless, all these investigations are usually carried out under steady state thermal conditions. Ultrashort laser induced thermal management and the ensuing thermionic control would add dynamical aspects to these measurements, which would potentially unravel new information boosting our current understanding on these topics in applied surface science. The methodology and the results presented in our study would additionally benefit the interpretation of the results from and the design of such experiments.

\section*{CRediT authorship contribution statement}
\textbf{Mousumi Upadhyay Kahaly}: Writing - original draft, Supervision, Conceptualization, Funding acquisition, Resources. \textbf{Saibabu Madas}: Writing - original draft, Software, Validation, Visualization, Formal analysis, Investigation. \textbf{Boris Mesits}: Software, Writing - review and editing, Validation, Investigation. \textbf{Subhendu Kahaly}: Supervision, Conceptualization, Methodology, Project administration, Writing - review and editing, Visualization, Funding acquisition.

\section*{Declaration of Competing Interest}
The authors confirm that they have no competing financial interests or personal relationships that could have influenced the findings presented in this paper.

\section*{Acknowledgements} 
The ELI ALPS project (GINOP-2.3.6-15-2015-00001) is supported by the European Union and it is co-financed by the European Regional Development Fund. This research has been supported by the IMPULSE project which receives funding from the European Union Framework Programme for Research and Innovation Horizon 2020 under grant agreement No 871161. SK, MUK and SM also acknowledges project No. 2019-2.1.13-T\'ET-IN-2020-00059, which has been implemented with support provided by the National Research, Development and Innovation Fund of Hungary, and financed under the 2019-2.1.13-T\'ET-IN funding scheme. SM would like to thank the ELI HPC administration for their support in providing computational resources.

%





\end{document}